\documentclass[]{aastex}

\usepackage{emulateapj5}
\usepackage{onecolfloat}
\usepackage{graphicx} 
\usepackage{fancyheadings} 
\usepackage{ulem}
\usepackage{rotating}
\usepackage{lscape}

\newcommand{\ndla}{121}
\newcommand{\nfeh}{8}

\newcommand{\ntot}{300}
\newcommand{\nbin}{5}

\newcommand{\lya}{Ly$\alpha$ }

\newcommand{\cm}[1]{\, {\rm cm^{#1}}}
\newcommand{\N}[1]{{N({\rm #1})}}

\newcommand{\sci}[1]{{\rm \; \times \; 10^{#1}}}

\newcommand{\tskip}{\tablevspace{1pt}}

\begin{document}

\twocolumn[%
\submitted{Submitted to Astrophysical Journal Letters: May 8, 2003}
\title{The Age-Metallicity Relation of the
Universe in Neutral Gas: The First 100 Damped \lya Systems}

\author{ Jason X. Prochaska\altaffilmark{1,2},
Eric Gawiser\altaffilmark{1,3,4,5}, 
Arthur M. Wolfe\altaffilmark{1,3}, 
Sandra Castro\altaffilmark{1,6,7},
S. G. Djorgovski\altaffilmark{1,6}}

\begin{abstract} 

We present accurate metallicity measurements for \ndla\ damped \lya systems
at $0.5 < z < 5$ including $\approx 50$ new measurements from our
recently published Echellette Spectrograph and Imager surveys.  This
dataset is analysed to determine the age-metallicity relation of 
neutral gas in the universe.  Contrary to previous datasets this sample 
shows statistically significant evolution in the mean metallicity. 
The best linear fit rate to metallicity vs.\ redshift is $-0.26 \pm 0.06$~dex
corresponding to approximately a factor of 2 every Gyr at $z=3$.  
The DLA continue to maintain a floor in metallicity of $\approx 1/700$ solar
independent of observational effects.  This metallicity
threshold limits the prevalence of primordial gas in high redshift
galaxies and stresses the correspondence between damped systems and
star formation (i.e.\ galaxy formation).  This floor is significantly offset
from the metallicity of the \lya forest and therefore we consider it to
be more related to active star formation within these galaxies than scenarios
of enrichment in the very early universe.
Finally, we comment on an apparent 'missing metals problem': the mean 
metallicity of the damped systems is $\approx 10\times$ lower than
the value expected from their observed star formation history.
This problem is evident in current theoretical treatments of chemical
evolution and galaxy formation; it may indicate a serious flaw in
our understanding of the interplay between star formation and metal production.

\keywords{galaxies: abundances --- 
galaxies: chemical evolution --- quasars : absorption lines }

\end{abstract}
]

\altaffiltext{1}{Visiting Astronomer, W.M. Keck Telescope.
The Keck Observatory is a joint facility of the University
of California, California Institute of Technology, and NASA.}
\altaffiltext{2}{UCO/Lick Observatory, University of California, Santa Cruz,
Santa Cruz, CA 95064}
\altaffiltext{3}{Department of Physics, and Center for Astrophysics and 
Space Sciences, University of California, San Diego, C--0424, La Jolla, 
CA 92093-0424}
\altaffiltext{4}{NSF Postdoctoral Fellow, Yale University, New Haven, CT, PO Box 208101, New Haven, CT  06520}
\altaffiltext{5}{Andes Prize Fellow, Universidad de Chile, Casilla 36-D, Santiago, Chile}
\altaffiltext{6}{Palomar Observatory, 105-24, California Institute of Technology, Pasadena, CA  91125}
\altaffiltext{7}{Infrared Processing and Analysis Center, 100-22, California 
Institute of Technology, Pasadena, CA  91125}

\pagestyle{fancyplain}
\lhead[\fancyplain{}{\thepage}]{\fancyplain{}{PROCHASKA ET AL.}}
\rhead[\fancyplain{}{The Age-Metallicity Relation of the
Universe in Neutral Gas}]{\fancyplain{}{\thepage}}
\setlength{\headrulewidth=0pt}
\cfoot{}

\section{INTRODUCTION}
\label{sec-intro}

For the past decade researchers have observed the damped \lya
systems (DLA) -- quasar absorption line systems with H\,I column density
$\N{HI} > 2 \sci{20} \cm{-2}$ -- to trace cosmological properties
of neutral gas in the early universe.
With moderate resolution spectroscopy, for example, observers have taken
a census of the H\,I mass density $\Omega_{gas}$ from $z = 0$ to 5 and found
damped \lya systems to comprise most of the neutral gas out to at least $z=4$
\citep{wolfe86,lzwt95,wolfe95,storrie00,rao00,peroux01,djg03}.
By combining these observations with higher resolution spectroscopy of metal-line
transitions, one tracks
the metal enrichment of the universe in neutral gas
\citep{pettini94,pettini99,pw00}.
If the individual DLA metallicity measurements are weighted by their
corresponding H\,I column densities, the resulting mean represents
a cosmological quantity: $\Omega_{metals}/\Omega_{gas}$, which equals
the mass-weighted metallicity $<Z>$ of neutral gas
in the universe \citep{lzwt95}.  
Aside from selection biases, 
this statistic is independent of any physical property (e.g.\ mass, morphology)
of the damped systems surveyed and, therefore, it presents a fundamental
test for theories of chemical evolution \citep[e.g.][]{pei95,somerville01}.
In turn, these observations constrain the star formation history of the
universe and help describe the interplay between nucleosynthesis and
gas enrichment in high redshift galaxies.

Because $<Z>$ is a $\N{HI}$-weighted measure, the uncertainty in this
statistic is dominated by the
damped systems with the highest product of metallicity and H\,I column
density.  This is analogous to measurements of $\Omega_{gas}$ where damped
systems with the largest $\N{HI}$ dominate the uncertainty\footnote{This effect
is less severe at $z>4$ where there are fewer DLA with 
$\N{HI} > 10^{21} \cm{-2}$ \citep[see][]{peroux01}}.
For this reason, previous metallicity samples 
were susceptible to severe sample variance, 
systematic error,  and potential outliers.
With the goal of reducing the effect of small number statistics, 
we initiated a program
\citep{pgw01} with the Echellette Spectrograph and Imager 
\citep[ESI;][]{sheinis02} to rapidly increase the sample of 
high $z$ damped systems with accurate metallicity measurements.
In comparison with previous echelle observations,
this instrument and observing strategy have led to a nearly 10x
increase in efficiency.  In roughly 5 nights observing time,
we have doubled the number of $z>1.5$ damped systems and nearly 
quadrupled the systems at $z>3$ \citep{p03a,pro03b}. 
In this Letter, we report the principal results on chemical 
evolution from our ESI surveys of the damped \lya systems.
Combining these new measurements
with $\approx$ 50 damped systems drawn from the literature, our analysis
includes a sample of over 100 damped \lya systems from $z= 0.5$ to 5.

\begin{table*}\footnotesize
\begin{center}
\caption{ 
{\sc SUMMARY\label{tab:sum}}}
\begin{tabular}{lcccccccc}
\tableline
\tableline
QSO
& $z_{abs}$
& $\N{HI}$ 
& f$_{[M/H]}^a$
& [M/H]
& f$_{[Fe/H]}^b$
& [Fe/H]
& Instr
& Ref \\
\tableline\tskip
Q0235+1615  &0.526&21.80$^{+0.100}_{-0.100}$& 2&$-0.22\pm 0.15$& 0&$ 0.00\pm 0.00$&1       \\  
Q1622+238   &0.656&20.36$^{+0.100}_{-0.100}$& 4&$-0.87\pm 0.25$& 1&$-1.27\pm 0.15$&3,4     \\  
Q1122-168   &0.682&20.45$^{+0.050}_{-0.050}$& 4&$-1.00\pm 0.15$& 1&$-1.40\pm 0.05$&5       \\  
Q1328+307   &0.692&21.25$^{+0.060}_{-0.060}$& 2&$-1.81\pm 0.09$& 1&$-1.81\pm 0.08$&6       \\  
Q0454+039   &0.860&20.69$^{+0.060}_{-0.060}$& 1&$-0.80\pm 0.15$& 1&$-1.02\pm 0.11$&2,7     \\  
Q0302-223   &1.009&20.36$^{+0.110}_{-0.110}$& 1&$-0.74\pm 0.12$& 1&$-1.19\pm 0.12$&2       \\  
Q0948+43    &1.233&21.50$^{+0.100}_{-0.100}$& 2&$-0.99\pm 0.10$& 1&$-1.43\pm 0.10$&10      \\  
Q0935+417   &1.373&20.52$^{+0.100}_{-0.100}$& 2&$-1.21\pm 0.13$& 1&$-1.21\pm 0.13$&11      \\  
Q1354+258   &1.420&21.54$^{+0.060}_{-0.060}$& 1&$-1.74\pm 0.13$& 1&$-2.03\pm 0.08$&9       \\  
Q1104-18    &1.661&20.80$^{+0.100}_{-0.100}$& 1&$-1.04\pm 0.10$& 1&$-1.48\pm 0.10$&12      \\  
Q1331+17    &1.776&21.18$^{+0.041}_{-0.041}$& 1&$-1.45\pm 0.04$& 1&$-2.06\pm 0.04$&13,14   \\  
Q2230+02    &1.864&20.85$^{+0.084}_{-0.084}$& 1&$-0.76\pm 0.09$& 1&$-1.17\pm 0.09$&13,14   \\  
Q1210+17    &1.892&20.60$^{+0.100}_{-0.100}$& 1&$-0.88\pm 0.10$& 1&$-1.15\pm 0.12$&14      \\  
Q2206-19    &1.920&20.65$^{+0.071}_{-0.071}$& 1&$-0.42\pm 0.07$& 1&$-0.86\pm 0.07$&14,15   \\  
Q1157+014   &1.944&21.80$^{+0.100}_{-0.100}$& 2&$-1.36\pm 0.12$& 1&$-1.81\pm 0.11$&36      \\  
Q0551-366   &1.962&20.50$^{+0.080}_{-0.080}$& 1&$-0.44\pm 0.10$& 1&$-0.95\pm 0.10$&17      \\  
Q0013-004   &1.973&20.83$^{+0.070}_{-0.070}$& 1&$-0.96\pm 0.08$& 1&$-1.52\pm 0.08$&37      \\  
Q1215+33    &1.999&20.95$^{+0.067}_{-0.067}$& 1&$-1.48\pm 0.07$& 1&$-1.70\pm 0.09$&13,14   \\  
Q0010-002   &2.025&20.80$^{+0.100}_{-0.100}$& 2&$-1.20\pm 0.12$& 1&$-1.33\pm 0.11$&36      \\  
Q0458-02    &2.040&21.65$^{+0.090}_{-0.090}$& 2&$-1.19\pm 0.09$& 1&$-1.77\pm 0.10$&13,14   \\  
Q2231-002   &2.066&20.56$^{+0.100}_{-0.100}$& 1&$-0.88\pm 0.10$& 1&$-1.40\pm 0.12$&13,14   \\  
Q2206-19    &2.076&20.43$^{+0.060}_{-0.060}$& 1&$-2.31\pm 0.07$& 1&$-2.61\pm 0.06$&14,15   \\  
Q2359-02    &2.095&20.70$^{+0.100}_{-0.100}$& 1&$-0.78\pm 0.10$& 1&$-1.66\pm 0.10$&13,14   \\  
Q0528-2505  &2.141&20.70$^{+0.080}_{-0.080}$& 1&$-1.00\pm 0.09$& 1&$-1.26\pm 0.36$&7       \\  
Q0149+33    &2.141&20.50$^{+0.100}_{-0.100}$& 1&$-1.49\pm 0.11$& 1&$-1.77\pm 0.10$&13,14   \\  
Q2359-02    &2.154&20.30$^{+0.100}_{-0.100}$& 1&$-1.58\pm 0.10$& 1&$-1.88\pm 0.10$&13,14   \\  
Q2348-14    &2.279&20.56$^{+0.075}_{-0.075}$& 1&$-1.92\pm 0.08$& 1&$-2.24\pm 0.08$&13,14,18\\  
Q0216+08    &2.293&20.45$^{+0.160}_{-0.160}$& 1&$-0.56\pm 0.17$& 1&$-1.06\pm 0.18$&7       \\  
PH957       &2.309&21.37$^{+0.080}_{-0.080}$& 1&$-1.46\pm 0.08$& 1&$-1.90\pm 0.09$&14,19,20\\  
Q1232+08    &2.337&20.90$^{+0.100}_{-0.100}$& 1&$-1.22\pm 0.15$& 1&$-1.72\pm 0.13$&16      \\  
HE2243-6031 &2.330&20.67$^{+0.020}_{-0.020}$& 1&$-0.87\pm 0.03$& 1&$-1.25\pm 0.02$&21      \\  
Q0841+12    &2.375&20.95$^{+0.087}_{-0.087}$& 1&$-1.27\pm 0.09$& 4&$-1.78\pm 0.09$&13,14   \\  
Q0102-190   &2.370&20.85$^{+0.100}_{-0.100}$& 1&$-1.81\pm 0.11$& 1&$-1.89\pm 0.13$&36      \\  
Q2348-01    &2.426&20.50$^{+0.100}_{-0.100}$& 1&$-0.70\pm 0.10$& 1&$-1.39\pm 0.10$&36      \\  
Q2343+12    &2.431&20.34$^{+0.100}_{-0.100}$& 1&$-0.54\pm 0.10$& 1&$-1.20\pm 0.10$&36      \\  
Q0112-306   &2.418&20.37$^{+0.080}_{-0.080}$& 1&$-2.32\pm 0.10$& 1&$-2.52\pm 0.10$&13,14   \\  
Q0112+029   &2.423&20.78$^{+0.080}_{-0.080}$& 1&$-1.29\pm 0.11$& 1&$-1.46\pm 0.10$&22      \\  
Q1409+095   &2.456&20.54$^{+0.100}_{-0.100}$& 1&$-2.02\pm 0.10$& 1&$-2.30\pm 0.10$&23      \\  
Q0201+36    &2.463&20.38$^{+0.045}_{-0.045}$& 1&$-0.41\pm 0.05$& 1&$-0.87\pm 0.04$&14,24   \\  
Q0836+11    &2.465&20.58$^{+0.100}_{-0.100}$& 1&$-1.15\pm 0.11$& 1&$-1.40\pm 0.10$&14      \\  
Q1223+17    &2.466&21.50$^{+0.100}_{-0.100}$& 1&$-1.59\pm 0.10$& 1&$-1.84\pm 0.10$&14,25   \\  
Q0841+12    &2.476&20.78$^{+0.097}_{-0.097}$& 3&$-1.62\pm 0.22$& 1&$-1.75\pm 0.11$&13,14   \\  
Q1451+123   &2.469&20.39$^{+0.100}_{-0.100}$& 1&$-2.13\pm 0.14$& 1&$-2.46\pm 0.11$&36      \\  
Q2344+12    &2.538&20.36$^{+0.100}_{-0.100}$& 1&$-1.74\pm 0.10$& 1&$-1.83\pm 0.10$&7,14    \\  
Q0405-443   &2.550&21.00$^{+0.150}_{-0.200}$& 2&$-1.64\pm 0.29$& 1&$-1.76\pm 0.23$&36      \\  
Q1502+4837  &2.570&20.30$^{+0.150}_{-0.150}$& 1&$-1.62\pm 0.17$& 1&$-1.65\pm 0.19$&26      \\  
Q1209+0919  &2.584&21.40$^{+0.100}_{-0.100}$& 2&$-1.09\pm 0.11$& 1&$-1.68\pm 0.11$&26      \\  
Q0405-443   &2.595&20.90$^{+0.100}_{-0.100}$& 1&$-0.96\pm 0.10$& 1&$-1.33\pm 0.10$&36      \\  
Q2348-01    &2.615&21.30$^{+0.100}_{-0.100}$& 1&$-1.97\pm 0.12$& 1&$-2.23\pm 0.13$&13,14   \\  
FJ0812+32   &2.626&21.35$^{+0.100}_{-0.100}$& 1&$-0.96\pm 0.11$& 1&$-1.74\pm 0.10$&26,27   \\  
Q1759+75    &2.625&20.76$^{+0.007}_{-0.007}$& 1&$-0.79\pm 0.01$& 1&$-1.18\pm 0.01$&13,14,28\\  
Q0058-292   &2.671&21.10$^{+0.100}_{-0.100}$& 1&$-1.44\pm 0.13$& 1&$-1.86\pm 0.12$&36      \\  
CTQ460      &2.777&21.00$^{+0.100}_{-0.100}$& 1&$-1.41\pm 0.10$& 1&$-1.82\pm 0.10$&26      \\  
PKS1354-17  &2.780&20.30$^{+0.150}_{-0.150}$& 1&$-1.88\pm 0.16$& 1&$-2.43\pm 0.17$&26      \\  
HS1132+2243 &2.783&21.00$^{+0.070}_{-0.070}$& 1&$-2.07\pm 0.15$& 1&$-2.48\pm 0.10$&26      \\  
PSS1253-0228&2.783&21.85$^{+0.200}_{-0.200}$& 2&$-1.75\pm 0.21$& 1&$-1.99\pm 0.20$&26      \\  
Q1337+11    &2.795&20.95$^{+0.100}_{-0.100}$& 1&$-1.79\pm 0.15$& 1&$-2.39\pm 0.10$&26      \\  
Q1008+36    &2.799&20.70$^{+0.050}_{-0.050}$& 1&$-1.81\pm 0.05$& 3&$-1.11\pm 0.05$&14      \\  
\tableline
\end{tabular}
\end{center}
\end{table*}

\begin{table*}\footnotesize
\begin{center}
\begin{tabular}{lcccccccc}
& & & & Table 1 -- cont \\
\tableline\tskip
Q0135-273   &2.800&20.90$^{+0.100}_{-0.100}$& 1&$-1.47\pm 0.13$& 1&$-1.65\pm 0.17$&36      \\  
Q1425+6039  &2.827&20.30$^{+0.040}_{-0.040}$& 4&$-0.93\pm 0.14$& 1&$-1.33\pm 0.04$&7,14    \\  
Q2138-444   &2.852&20.80$^{+0.080}_{-0.080}$& 2&$-1.52\pm 0.13$& 1&$-1.72\pm 0.10$&36      \\  
Q2342+34    &2.908&21.10$^{+0.100}_{-0.100}$& 1&$-1.19\pm 0.10$& 1&$-1.62\pm 0.12$&26      \\  
BQ1021+3001 &2.949&20.70$^{+0.100}_{-0.100}$& 1&$-2.17\pm 0.10$& 1&$-2.32\pm 0.10$&26      \\  
BRJ0426-2202&2.983&21.50$^{+0.150}_{-0.150}$& 4&$-2.45\pm 0.26$& 1&$-2.85\pm 0.16$&26      \\  
HS0741+4741 &3.017&20.48$^{+0.100}_{-0.100}$& 1&$-1.69\pm 0.10$& 1&$-1.93\pm 0.10$&14      \\  
Q0347-38    &3.025&20.63$^{+0.005}_{-0.005}$& 1&$-1.17\pm 0.03$& 1&$-1.62\pm 0.01$&13,14   \\  
FJ2334-09   &3.057&20.45$^{+0.100}_{-0.100}$& 1&$-1.15\pm 0.12$& 1&$-1.63\pm 0.10$&26      \\  
Q0336-01    &3.062&21.20$^{+0.100}_{-0.100}$& 1&$-1.41\pm 0.10$& 1&$-1.79\pm 0.10$&14      \\  
PSS0808+52  &3.113&20.65$^{+0.070}_{-0.070}$& 1&$-1.61\pm 0.14$& 1&$-1.98\pm 0.08$&26,29   \\  
Q2223+20    &3.119&20.30$^{+0.100}_{-0.100}$& 1&$-2.22\pm 0.11$& 1&$-2.43\pm 0.11$&26      \\  
PSS1535+2943&3.202&20.65$^{+0.150}_{-0.150}$& 3&$-1.00\pm 0.30$&11&$-1.25\pm 0.30$&30      \\  
PSS2344+0342&3.219&21.35$^{+0.070}_{-0.070}$& 3&$-1.90\pm 0.15$& 1&$-1.62\pm 0.12$&26      \\  
PSS1506+5220&3.224&20.67$^{+0.070}_{-0.070}$& 1&$-2.35\pm 0.07$& 1&$-2.46\pm 0.08$&26      \\  
PSS2315+0921&3.219&21.35$^{+0.150}_{-0.150}$& 5&$-1.68\pm 0.31$&11&$-2.08\pm 0.21$&30      \\  
Q0930+28    &3.235&20.30$^{+0.100}_{-0.100}$& 1&$-1.97\pm 0.10$& 1&$-2.10\pm 0.10$&14      \\  
J0255+00    &3.253&20.70$^{+0.100}_{-0.100}$& 1&$-0.94\pm 0.11$& 1&$-1.44\pm 0.10$&14      \\  
PSS1432+39  &3.272&21.25$^{+0.100}_{-0.100}$& 1&$-1.14\pm 0.11$& 4&$-1.85\pm 0.14$&26,29   \\  
PSS0957+33  &3.280&20.45$^{+0.080}_{-0.080}$& 1&$-1.13\pm 0.10$& 1&$-1.58\pm 0.08$&14,26,29\\  
PSS2155+1358&3.316&20.55$^{+0.150}_{-0.150}$& 1&$-1.26\pm 0.17$&13&$-1.65\pm 0.15$&26      \\  
Q1055+46    &3.317&20.34$^{+0.100}_{-0.100}$& 1&$-1.65\pm 0.15$& 1&$-1.87\pm 0.10$&22      \\  
PSS1715+3809&3.341&21.05$^{+0.150}_{-0.100}$& 4&$-2.41\pm 0.26$& 1&$-2.81\pm 0.15$&30      \\  
BR1117-1329 &3.350&20.84$^{+0.100}_{-0.100}$& 1&$-1.27\pm 0.13$& 1&$-1.51\pm 0.10$&31      \\  
PSS1802+5616&3.391&20.30$^{+0.100}_{-0.100}$& 3&$-1.43\pm 0.15$& 1&$-1.54\pm 0.11$&30      \\  
Q0000-2619  &3.390&21.41$^{+0.080}_{-0.080}$& 1&$-1.91\pm 0.08$& 1&$-2.16\pm 0.09$&7,14,32 \\  
Q0201+11    &3.387&21.26$^{+0.100}_{-0.100}$& 1&$-1.25\pm 0.15$& 1&$-1.41\pm 0.11$&33      \\  
PC0953+47   &3.404&21.15$^{+0.150}_{-0.150}$& 3&$-1.82\pm 0.28$&13&$-1.89\pm 0.29$&26      \\  
FJ0747+2739 &3.423&20.85$^{+0.050}_{-0.050}$& 3&$-1.68\pm 0.20$&11&$-1.78\pm 0.14$&26      \\  
PSS2315+0921&3.425&21.10$^{+0.200}_{-0.200}$& 1&$-1.51\pm 0.21$&11&$-1.79\pm 0.17$&30      \\  
BR0019-15   &3.439&20.92$^{+0.100}_{-0.100}$& 1&$-1.06\pm 0.11$& 4&$-1.59\pm 0.11$&13,14   \\  
PSS0007+2417&3.496&21.10$^{+0.100}_{-0.100}$& 1&$-1.58\pm 0.11$& 4&$-1.92\pm 0.11$&30      \\  
PSS1802+5616&3.554&20.50$^{+0.100}_{-0.100}$& 4&$-1.52\pm 0.22$& 1&$-1.93\pm 0.12$&30      \\  
BRI1108-07  &3.608&20.50$^{+0.100}_{-0.100}$& 1&$-1.80\pm 0.10$& 1&$-2.12\pm 0.10$&13,14   \\  
PSS0209+0517&3.667&20.45$^{+0.100}_{-0.100}$& 3&$-1.73\pm 0.17$& 1&$-2.31\pm 0.11$&26      \\  
PSS2323+2758&3.684&20.95$^{+0.100}_{-0.100}$& 1&$-2.59\pm 0.10$& 1&$-3.13\pm 0.16$&26      \\  
PSS1248+31  &3.696&20.63$^{+0.070}_{-0.070}$& 1&$-1.80\pm 0.07$& 1&$-2.24\pm 0.08$&26,29   \\  
PSS0133+0400&3.693&20.70$^{+0.100}_{-0.150}$& 3&$-1.90\pm 0.19$& 1&$-2.69\pm 0.12$&26      \\  
PSS0007+2417&3.705&20.55$^{+0.150}_{-0.150}$& 3&$-1.50\pm 0.24$&11&$-1.64\pm 0.20$&30      \\  
PSS1723+2243&3.695&20.50$^{+0.150}_{-0.150}$& 3&$-0.61\pm 0.15$&13&$-1.16\pm 0.26$&26      \\  
SDSS0127-00 &3.727&21.15$^{+0.100}_{-0.100}$& 3&$-2.18\pm 0.23$& 0&$ 0.00\pm 0.00$&26      \\  
BRI1346-03  &3.736&20.72$^{+0.100}_{-0.100}$& 1&$-2.33\pm 0.10$& 6&$-2.63\pm 0.10$&13,14   \\  
PSS0133+0400&3.774&20.55$^{+0.100}_{-0.150}$& 1&$-0.64\pm 0.11$& 4&$-0.92\pm 0.10$&26      \\  
PSS1802+5616&3.762&20.55$^{+0.150}_{-0.150}$& 3&$-1.55\pm 0.19$&25&$-1.82\pm 0.26$&30      \\  
PSS0134+3317&3.761&20.85$^{+0.050}_{-0.100}$& 4&$-2.33\pm 0.20$& 6&$-2.73\pm 0.10$&26      \\  
PSS1535+2943&3.761&20.40$^{+0.150}_{-0.150}$& 1&$-2.02\pm 0.16$& 6&$-2.33\pm 0.16$&30      \\  
PSS1802+5616&3.811&20.35$^{+0.200}_{-0.200}$& 1&$-2.04\pm 0.22$& 1&$-2.19\pm 0.23$&30      \\  
PSS0007+2417&3.838&20.85$^{+0.150}_{-0.150}$& 3&$-2.19\pm 0.22$& 1&$-2.44\pm 0.15$&30      \\  
PSS0209+0517&3.864&20.55$^{+0.100}_{-0.100}$& 1&$-2.65\pm 0.11$& 6&$-2.89\pm 0.11$&26      \\  
BR0951-04   &3.857&20.60$^{+0.100}_{-0.100}$& 4&$-1.60\pm 0.22$& 1&$-2.00\pm 0.12$&13,14   \\  
PC0953+47   &3.891&21.20$^{+0.100}_{-0.100}$& 3&$-1.50\pm 0.15$& 4&$-1.80\pm 0.11$&26      \\  
FJ0747+2739 &3.900&20.50$^{+0.100}_{-0.100}$& 1&$-2.03\pm 0.10$& 6&$-2.53\pm 0.10$&26      \\  
J0255+00    &3.915&21.30$^{+0.050}_{-0.050}$& 1&$-1.78\pm 0.05$& 1&$-2.05\pm 0.10$&14      \\  
\tableline
\end{tabular}
\end{center}
\end{table*}

\begin{table*}\footnotesize
\begin{center}
\begin{tabular}{lcccccccc}
& & & & Table 1 -- cont \\
\tableline\tskip
BRI0952-01  &4.024&20.55$^{+0.100}_{-0.100}$& 4&$-1.46\pm 0.23$& 1&$-1.86\pm 0.13$&13,14   \\  
BR2237-0607 &4.080&20.52$^{+0.110}_{-0.110}$& 1&$-1.87\pm 0.11$& 1&$-2.14\pm 0.17$&7       \\  
PSS0957+33  &4.180&20.70$^{+0.100}_{-0.100}$& 1&$-1.70\pm 0.10$& 1&$-2.07\pm 0.11$&26,29   \\  
BR0951-04   &4.203&20.40$^{+0.100}_{-0.100}$& 1&$-2.62\pm 0.10$& 3&$-2.59\pm 0.10$&13,14   \\  
PSS1443+27  &4.224&20.80$^{+0.100}_{-0.100}$& 4&$-0.70\pm 0.21$& 1&$-1.10\pm 0.11$&14,25   \\  
PC0953+47   &4.244&20.90$^{+0.150}_{-0.150}$& 1&$-2.23\pm 0.15$& 1&$-2.50\pm 0.17$&26      \\  
PSS2241+1352&4.282&21.15$^{+0.100}_{-0.100}$& 1&$-1.77\pm 0.10$&11&$-1.90\pm 0.11$&26      \\  
BR1202-07   &4.383&20.60$^{+0.140}_{-0.140}$& 1&$-1.81\pm 0.14$& 1&$-2.19\pm 0.19$&7       \\  
J0307-4945  &4.468&20.67$^{+0.090}_{-0.090}$& 1&$-1.55\pm 0.12$& 1&$-1.96\pm 0.23$&34      \\  
SDSS1737+582&4.743&20.65$^{+0.150}_{-0.150}$& 1&$-1.88\pm 0.16$& 1&$-2.39\pm 0.17$&35      \\  
\tableline
\end{tabular}
\end{center}
$^a$1=Si,S, or O; 2=Zn; 3=$\alpha$-element + Zn limits;
4=Fe+0.4; 5=Fe limits + 0.4\\
$^b$1=Fe; 4=Ni-0.1; 5=Cr-0.2; 6=Al; 11-16=Fe,Ni,Cr,Al limits \\
Key to References -- 
1: \cite{jnk03}; 
2: \cite{ptt00}; 
3: \cite{churc00}; 
4: \cite{rao00}; 
5: \cite{ledoux02a}; 
6: \cite{boisse98}; 
7: \cite{lu96}; 
9: \cite{pettini99}; 
10: \cite{pro03c}; 
11: \cite{meyer95}; 
12: \cite{lopez99}; 
13: \cite{pw99}; 
14: \cite{pro01}; 
15: \cite{pw97}; 
16: \cite{srianand00}; 
17: \cite{ledoux02b}; 
18: \cite{pettini95}; 
19: \cite{wolfe94}; 
20: \cite{dessauges03}; 
21: \cite{lopez02}; 
22: \cite{lu99}; 
23: \cite{pettini02}; 
24: \cite{pw96}; 
25: \cite{pw00}; 
26: \cite{pro03a}; 
27: \cite{phw03}; 
28: \cite{pro02}; 
29: \cite{pgw01}; 
30: \cite{pro03b}; 
31: \cite{peroux02}; 
32: \cite{molaro00}; 
33: \cite{ellison01}; 
34: \cite{dessauges01}; 
35: \cite{songaila02} 
36: \cite{ledoux03} 
37: \cite{petit02} 
\end{table*}

\clearpage

\begin{figure*}
\begin{center}
\includegraphics[height=5.3in, width=3.8in,angle=-90]{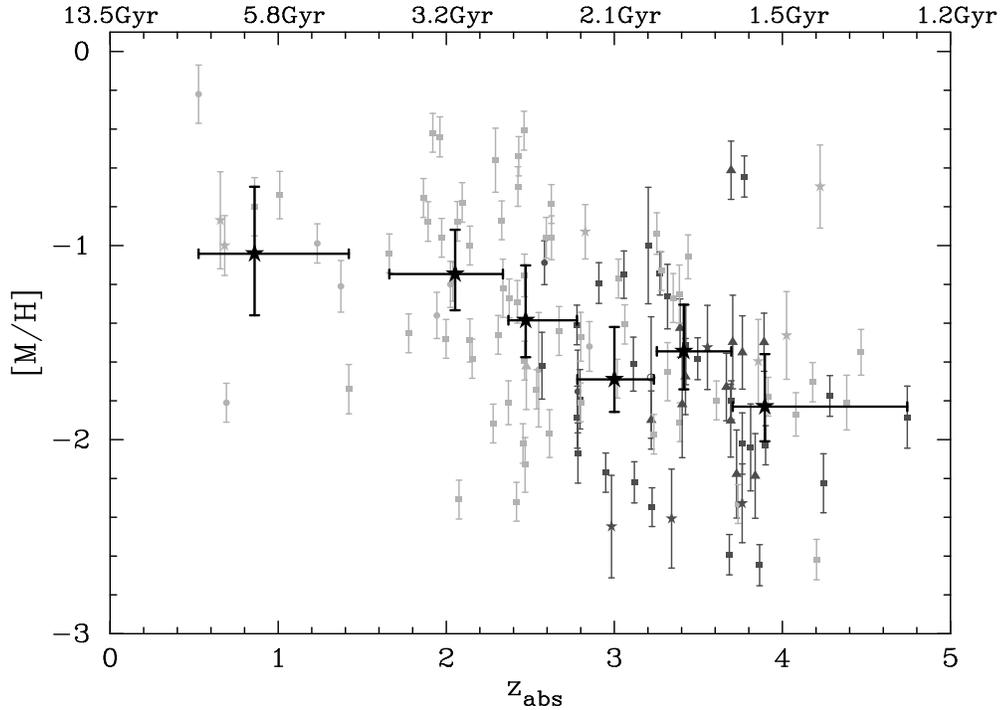}
\figcaption{The [M/H], $z_{abs}$ values for the \ndla\ damped systems comprising
our complete sample.  The darker points indicate the new values from our 
ESI surveys while the lighter data points are all taken from the literature
(primarily HIRES and UVES observations).  The symbols for the points indicate
the origin of the [M/H] values: $\alpha$-element measurement (filled squares);
Zn measurement (filled circle); $\alpha$+Zn limits (filled triangle); 
Fe measurement + 0.4~dex (filled star); Fe limits + 0.4~dex (open circle).
A trend of lower metallicity at higher redshift is clearly evident in the
figure.  We evaluate this trend by measuring the unweighted mean metallicity
in the 6 bins (solid star points with error bars).  A least-squares fit to
these data points yields a best fit slope of $m = -0.26 \pm 0.06$~dex.
The error bars plotted on the binned data represent 95$\%$~c.l. derived from a
bootstrap error analysis.
}
\label{fig:unweight}
\end{center}
\end{figure*}

\section{THE SAMPLE}
\label{sec:sample}

At present over \ntot\ damped \lya systems have been identified 
in the literature \citep{curran02}.  Of these, approximately half provide a 
metallicity estimate.
Unfortunately, these metallicity measurements are derived from a
variety of telescopes 
with a range of instrumentation and therefore form a heterogeneous
sample of data (e.g.\ varying resolution, wavelength coverage, and S/N).
In this Letter, we primarily restrict\footnote{
The exceptions are the $z<1.5$ damped systems which include an estimate
from X-Ray observations (Junkkarinen et al.\ 2003, in preparation).} 
our analysis to damped \lya systems observed
on the current generation of large telescopes (e.g.\ Keck and the VLT)
with high-resolution ($R > 5000$), high S/N ($>15$ per pixel) spectra.
Although this is a subjective observational criterion,
there are many cases of understated statistical error and
under-appreciated systematic error in the literature for data of
poorer quality.  We also limit the analysis to systems satisfying
the strict $\N{HI} \geq 2 \sci{20} \cm{-2}$ criterion.  
This practice facilitates comparisons with
statistical surveys of the damped systems and simplifies comparisons
with theoretical models.

Table~\ref{tab:sum} lists the name, $z_{abs}$, $\N{HI}$, metallicity
[M/H], Fe abundance [Fe/H], and reference for the \ndla\ DLA comprising
the complete sample.  In columns 4 and 6 we list two `flags' which describe
the derivation of the [M/H] and [Fe/H] values.  The latter are
determined primarily from Fe\,II transitions or when 
necessary Cr\,II, Ni\,II, or Al\,II transitions offset 
by their typical [X/Fe] value
(e.g.\ Prochaska \& Wolfe; hereafter, PW02).  
When possible, we adopt an [M/H] value
based on an observed $\alpha$-element (e.g.\ Si, S) or Zn.
These elements are mildly or non-refractory and should exhibit minimal
depletion in the damped \lya systems.  Furthermore, these elements
exhibit solar relative abundances with few exceptions (PW02).
In cases where there are only limits reported to these abundances and
the limits span an interval less than 0.4~dex (e.g.\ $\pm 0.2$~dex),
we adopt the central value and an error encompassing the two limits.
As a last resort (\nfeh\ cases), we adopt an [M/H] value calculated from
[Fe/H] assuming an offset of 0.4~dex. 
We consider this robust because the median and mean [$\alpha$/Fe]
and [Zn/Fe] values for DLA at $z>2$ are all close to 0.4~dex (PW02).

We present the full set of [M/H] values as a function of $z_{abs}$ in
Figure~\ref{fig:unweight}.  The dark, unbinned points identify the data 
drawn from our recent ESI surveys while the light unbinned points are
drawn from our echelle measurements and those of others
reported in the literature.
The error bars represent the statistical uncertainty in these
measurements.  For the majority of these observations, this error is dominated
by the uncertainty in $\N{HI}$ measurements, an uncertainty that is
typically not rigorously derived.  The quoted values tend to overestimate
the statistical error and underestimate the systematic error from line-blending
and continuum placement.
In the following analysis, we adopt a minimum error of $0.1$~dex for
all metallicity measurements.  It is important to emphasize, however, 
that none of the analysis in this Letter is sensitive to 
statistical error related to individual measurements.

\begin{figure*}
\begin{center}
\includegraphics[height=5.3in, width=3.8in,angle=-90]{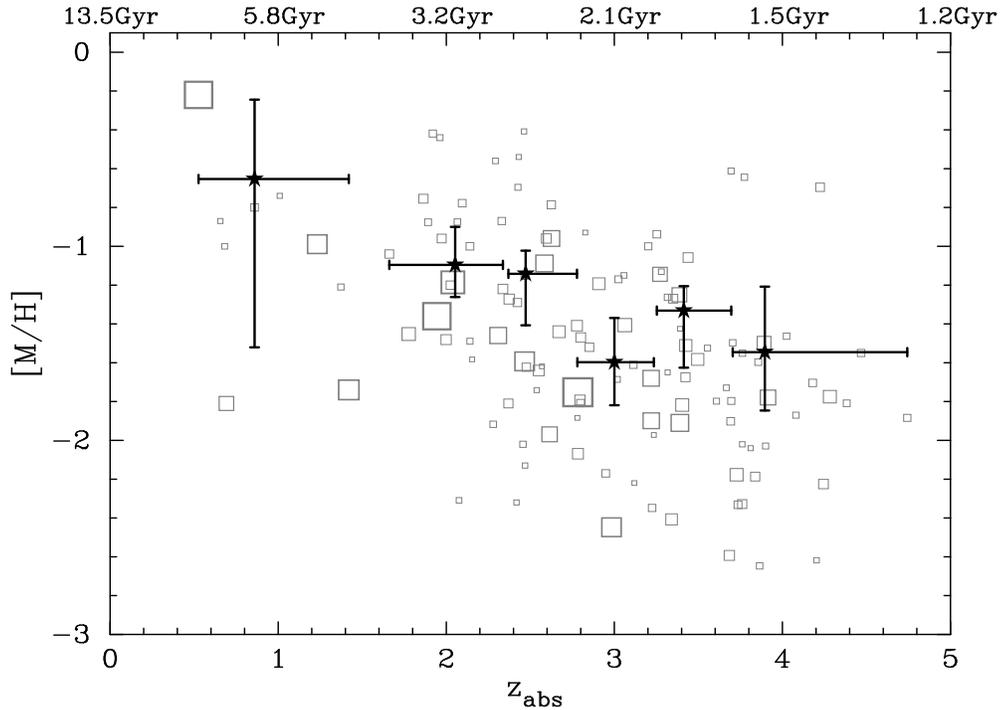}
\figcaption{The [M/H], $z_{abs}$ pairs are plotted as open squares where the area
of each square is scaled to the $\N{HI}$ value of the damped system.  This
aids the eye in determining which sightlines dominate the $\N{HI}$-weighted,
cosmic mean metallicity.  The cosmic metallicity $<Z>$ is plotted for 6 bins
with 95$\%$c.l. uncertainties given by a bootstrap analysis.  
Similar to the unweighted mean,
we find a best fit slope of $m = -0.25 \pm 0.07$~dex assuming a linear solution
to the $<Z>$ vs.\ $z_{abs}$ values.
}
\label{fig:weight}
\end{center}
\end{figure*}

\section{ANALYSIS}
\label{sec:analysis}

In Figure~\ref{fig:unweight} we overplot the unweighted, 
logarithmic mean metallicity (i.e.\ the mean of [M/H] values)
in \nbin\ redshift bins at $z>1.5$ defined to have equal numbers of DLA and
one redshift bin with $z=[0.5,1.5]$ covering the damped systems
with \lya profiles at $\lambda_{obs} < 3100$\AA.
This statistic was chosen to best represent the evolution in the metallicity
of a `typical' damped system
as seen by eye in the log-linear plot in Figure~\ref{fig:unweight}.
The vertical error bars report {\it 95$\%$~c.l.\ uncertainty} in this mean value
as determined from a bootstrap error analysis. 
Meanwhile, the redshift marked for each bin refers to the median $z_{abs}$ value.
Even an eyeball analysis of the figure reveals a statistically 
significant evolution in the unweighted mean metallicity.  
Performing a least-squares linear fit to the binned data (which 
best accounts for the scatter in the measurements), we calculate
a slope $m=-0.26 \pm 0.06$~dex/$\Delta z$ and zero-point
[M/H]$_0 = -0.67 \pm 0.17$~dex.  
The results indicate that the
metallicity of the `average' galaxy is increasing with decreasing
redshift with an e-folding time of approximately 1 unit redshift, i.e., 
a 2$\times$ decrease per Gyr at $z \sim 3$.
This statistically significant evolution in the unweighted mean contrasts with our
previous samples, although \cite{vladilo00} suggested such a trend at $z<3$ in
a much smaller sample.
We note that the y-intercept of the best-fit line is nearly $4\sigma$ below
solar metallicity.
This may be an indication of small sampling at low $z$ or our assumption
that the logarithmic mean metallicity evolves linearly with redshift.
On the other hand, it is possible that a cross-section selected sample
of H\,I gas at $z=0$ \citep[e.g.][]{rosen03} would show a sub-solar 
metallicity due to the contributions from dwarf and low surface-brightness
galaxies.

The cosmological mean metallicity $<Z>$ 
was computed using 
$<Z>= \log [ \sum_i 10^{[M/H]_i} \N{HI}_i / \sum_i \N{HI}_i ]$ 
and is presented in Figure~\ref{fig:weight} in the same six bins.  
In the figure, the 
individual measurements are now represented by squares whose areas
scale with the $\N{HI}$ values of the DLA. 
Because the $<Z>$ statistic is dominated by the DLA with the largest 
$\N{HI}$ and [M/H] values, its measurement uncertainty will always be
dominated by sample variance as opposed to statistical error.
We estimate this variance through the bootstrap technique and the
vertical error bars refer to 95$\%$~c.l.\ on the mean.
The bootstrap technique assesses the sample variance under the 
assumption that the observed distribution is not severely different
from the true, underlying distribution.
If the data sample is too small, even a single outlier 
could significantly change the central value of the mean statistic
significantly beyond the error implied by a bootstrap analysis.
While this issue was a concern
in previous analyses,  the results at $z \approx 2-4$ are now
robust to all but the unlikeliest of outliers.  For example, the discovery of
a system with $\N{HI} = 10^{22} \cm{-2}$ and 1/3 solar metallicity
would only increase $<Z>$ by $\approx +0.2$~dex in the $z \sim 2$ bin or
$+0.25$~dex at $z \sim 3.5$ \citep[see also][]{pro03}.
Such an outlier would lie nearly a factor of 10 off the current $\N{HI}$, [M/H]
distribution and it would
necessitate an entire population of DLA separate from the 100 sightlines
presented here.  This population could only exist if the current sample
is significantly biased by selection effects (e.g.\ dust obscuration),
an assertion unsupported by any recent observational test
(e.g.\ Ellison et al.\ 2001; PW02). 

Performing a least-squares fit to the $<Z>$ values, we find 
$m = -0.25 \pm 0.07$ and $b = -0.61 \pm 0.21$~dex.  
Note that the inclusion of the $z<1.5$
bin has minimal impact on this analysis.  Somewhat to our surprise,
especially given the impression from a visual inspection of Figures~1 and 2,
the best fit slope to the $<Z>$ values is identical within statistical
uncertainties to the value derived
for the unweighted mean.  This marks the first demonstration of evolution in
$<Z>$ at greater than $3 \sigma$ confidence.
In this case, our analysis implies that the mean metallicity of the universe
in neutral gas is approximately doubling every billion years from
$z = 4$ to 2.

\section{DISCUSSION}

The history of metallicity in the universe provides a potentially powerful 
constraint on models of galaxy formation, and it is tempting to draw 
additional conclusions from the distribution of points in the figures.  
The scatter amongst the points in each bin does not appear to evolve 
with redshift, implying that at each epoch DLA represent a set of 
systems at various stages in their formation.  
The interpretation of this is complicated by the unknown breakdown
of the observed scatter into scatter amongst galaxies and scatter within
galaxies.  Sightlines through the DLA provide pinpricks of transverse
diameter $\lesssim 10$~pc through the absorbing system.  Although DLA
abundances show remarkable uniformity amongst multiple components along
these sightlines \citep{pro03a}, we do not know how well the metallicity
of the entire galaxy is sampled by these measurements.
These characteristics of an evolving mean with nearly constant scatter
would like pose a difficult challenge to scenarios which would describe
damped \lya systems as a transient phase in the formation of galaxies.

Another important characteristic of the observed distribution of 
[M/H],$z$ pairs is the areas of redshift-metallicity space 
that remain unoccupied. In particular, note the absence of any
DLA at [M/H] $<-3$ at all redshifts and the lack 
of DLA at [M/H]$>0$ at all redshifts.  The eye can identify a gradual 
increase in the maximum and minimum metallicity with redshift at roughly 
the same rate of evolution as that of the unweighted logarithmic mean, 
but it would be dangerous to interpret this since the max and min are 
sensitive to outliers.    
Regarding the lower limit to the DLA metallicities, it appears
possible that we will never identify a damped \lya system with [M/H]~$<-3$,
a value which significantly exceeds our detection limit.  
This lower bound has important implications for
the presence of primordial gas (zero metallicity) within these galaxies.  
If primordial gas with significant surface density and cross-section 
exists in high redshift galaxies, then it is 
always surrounded by metal-enriched gas yielding a mass-weighted metallicity
exceeding 1/1000 solar.  Alternatively, 
primordial gas may not exist in the neutral phase within high $z$ galaxies. 
It may be possible to distinguish between these scenarios by observing
the higher order Lyman lines to pursue metallicity measurements of 
individual DLA components.

This metallicity floor also limits the contribution to DLA from 
gas `clouds'
which are unrelated to galaxies (e.g.\ overdense regions associated
with large-scale structures).
The metallicity floor requires that this gas was enriched to [M/H]~$> -3$, 
presumably by a nearby galaxy.  To match the observed metallicities would
probably require fine-tuning, which argues against the likelihood that
DLA are anything but galaxies in the early universe.
We also emphasize that the lower bound to the metallicity of the DLA 
([M/H]~$\approx -2.6$) is significantly higher than the metallicities
typically attributed to the \lya forest \citep[e.g.][]{songaila01}.  
This suggests that the
lower bound to the DLA values may not be simply related to the physical
processes which have enriched the \lya forest (e.g.\ pollution from
Population~III stars).  We contend that the offset between the DLA metallicities
and the \lya forest indicates the enrichment of DLA gas at all metallicity
is dominated by metal production {\it within} these galaxies.
It is possible, however, that Pop~III scenarios would predict larger
pre-enrichment in regions of higher overdensity, perhaps even to the
lower bound observed for the DLA.

A key implication of our results comes to light in their 
comparison with the recent determination of DLA star formation rates 
\citep{wpg03}.   Although it is not yet possible to compare 
the metallicity and integrated star formation rates of individual DLA,
it {\it is} possible to compare the average DLA metallicity at 
$z=2.5$ with the integral under the cosmic star formation history at 
$z>2.5$ \citep{wgp03}.  The comparison illustrates a ``missing metals 
problem'' for the DLA: the integrated SFR's measured for the DLA 
combined with a standard
IMF and yield of heavy elements
imply 10$\times$ the mass density of metals observed in the
damped systems.  This conflict suggests scenarios where the metals are
ejected from the galaxy via supernovae feedback or where the metals are
sequestered within the star formation regions for significant timescales
(e.g.\ within a galactic bulge; Wolfe, Gawiser, \& Prochaska 2003).
It is also possible that the mean metallicities of the typical DLA
systems are not representative of those in the actual star forming regions.
Observations of absorption systems associated with GRBs, which probably do
probe the star forming regions themselves, suggest higher metallicities
which may be more typical of disk stellar populations (see, e.g., Mirabal
et al. 2002, Savaglio et al. 2003, or Djorgovski et al. 2003, and references
therein).  
A fundamental uncertainty in comparing the actively star forming regions
sampled by GRBs and that of the majority of the neutral gas in the universe
sampled by the DLA is our lack of understanding of the mixing efficiency
of metal enrichment.
More detailed modeling is needed before this issue can be
completely resolved.

This problem is evident in recent theoretical models of chemical evolution
which consider the mean metallicity of the universe in neutral gas 
\citep{pei99,somerville01,tissera01,mathlin01,cen02}.  These treatments
range from analytic chemical evolution scenarios 
(i.e.\ independent of galaxy formation models) 
to full-blown hydrodynamic, numerical simulations.
The analyses appear to offer a wide range of theoretical values
for the mean metallicity of the universe in neutral gas, yet the majority
of variation stems from two key factors: 
(1) the adopted (or predicted) star formation history; and
(2) the treatment of selection biases due to dust obscuration.
An appreciation of these two aspects can be obtained by following the
analysis of \cite{somerville01} who focused on the stellar properties
of the Lyman Break Galaxies.
In addition to discussing their predictions for the star formation history of the
early universe, the authors examined the mean metallicity in cold gas.
In all of their scenarios, the values exceeded the damped \lya observations
by at least a factor of 3 at all redshifts $z>2$. 
This discrepancy is a restatement of the ``missing metals problem'' 
described above.  At present, the star formation history implied by
various studies of high-$z$ galaxies (including DLA themselves)
produce too many metals in 
comparison with the DLA.  This is a universal characteristic of these
chemical evolution models, even in those scenarios which underpredict
the star formation history at $z>2$ \citep[e.g.][]{pei99,mathlin01}.

Several of the theoretical treatments, however, 
report successes in matching previous
DLA metallicity samples.  In all of these cases, the authors introduced
significant selection biases owing to the effects of dust obscuration. 
Undoubtedly, dust obscuration plays a role in observations of the damped
\lya systems, especially at lower redshift where the gas metallicity
and dust content are presumably highest.  At $z>2$, however, current
samples of DLA toward radio-selected quasars exhibit no significant
difference from optically-selected samples \citep{ellison01b}.  
Furthermore, PW02 found no dependence between the inferred dust
opacity of observed DLA sightlines and the quasar magnitude, contrary
to expectation in scenarios where dust obscuration is important.
The original claims of reddening by \cite{pei91} have not been confirmed
by larger, homogeneous quasar samples, e.g., \cite{outram01} found only a $2\sigma$
indication of reddening in their $z \lesssim 1$ 2dF quasar sample.
At present, we believe there is no compelling evidence for dust 
obscuration\footnote{We distinguish obscuration from depletion, since in DLA
there is strong evidence from relative abundance ratios 
(e.g.\ Pettini et al.\ 1994; PW02) for the depletion of refractory elements
onto grains.}
beyond its convenience as a possible solution to the missing metals problem.
Regarding this aspect, we emphasize that
no group has self-consistently matched the observed star formation history
of the early universe with the metallicity measurements observed for the
damped systems even when allowing for dust obscuration.  
Current theories of star formation and metal production in high redshift
galaxies are missing a vital aspect of the processes.

Looking toward the future, 
observations of 300 additional DLA are needed to bring the statistical
error at $z=2$ to 4 
down to the level associated with systematic effects such as 
differential depletion.
With the introduction of ESI and similar future
instrumentation, we expect this will be accomplished within the next decade.
Significant gains, meanwhile, can be made very quickly at
$z<1.5$ and $z>4$ owing to the small sample size.  
Unfortunately, measurements at $z<1.5$ are slowed by the 
requirement of ultraviolet
observations.  This is a particularly challenging aspect of
damped \lya research because this redshift range 
corresponds to over half the age of the current universe.
While the COS spectrograph on the Hubble Space Telescope 
will improve the situation, a comprehensive view of metal enrichment over
the past 8~Gyr will require the launch of a 4m-class ultraviolet telescope
\citep{pro02}.

\acknowledgments

The authors wish to recognize and acknowledge the very significant cultural
role and reverence that the summit of Mauna Kea has always had within the
indigenous Hawaiian community.  We are most fortunate to have the
opportunity to conduct observations from this mountain.
We acknowledge the Keck support staff for their efforts
in performing these observations.  
We would also like to thank M. Pettini for helpful comments.
We also acknowledge our collaborators who have allowed us to present
several measurements prior to publication.
E.G. is supported by 
Fundaci\'{o}n Andes and by an NSF Astronomy and Astrophysics Postdoctoral 
Fellowship under award AST-0201667.  
AMW is partially supported by NSF grants AST 0071257.
SGD acknowledges a partial support from the Ajax Foundation.


\begin{thebibliography}{}

\bibitem[Boiss\'e et al.\ (1998)]{boisse98}	
Boiss\'e, P., Le Brun, V., Bergeron, J., \&
Deharveng, J.-M.  1998, \aap, 333, 841

\bibitem[Cen et al.\ (2003)]{cen02} 		
Cen, R., Ostriker, J.P., Prochaska, J.X., \& Wolfe, A.M. 2003, \apj, 
submitted

\bibitem[Churchill et al.\ (2000)]{churc00}    
Churchill, C.W., Mellon, R.R., Charlton, J.C., 
Jannuzi, B.T., Kirhakos, S., Steidel, C.C., \& Schneider, D.P. 2000,
\apjs, 130, 91


\bibitem[Curran et al.\ (2002)]{curran02}       	
Curran, S.J., Webb, J.K., Murphy, M.T., Bandiera, R., 
Corbelli, E., \& Flambaum, V.V. 2002, Publ. Astron. Soc. Austral.,
19, 455

\bibitem[Dessauges-Zavadsky et al.\ (2003)]{dessauges03}
Dessauges-Zavadsky et al.\ 2003, in preparation

\bibitem[Dessauges-Zavadsky et al.\ (2001)]{dessauges01}	
Dessauges-Zavadsky, M., D'Odorico, S., McMahon, R.G., 
Molaro, P., Ledoux, C., P${\rm \acute e}$roux, C., \&
Storrie-Lombardi, L.J. 2001, \aap, 370, 426

\bibitem[Djorgovski et al.\ (2003a)]{djg03a}		
Djorgovski, S.G., et al. 2003, in: Gamma-Ray Bursts in the Afterglow Era:
    3rd Workshop, ASPCS, in press (astro-ph/0302004)

\bibitem[Djorgovski et al.\ (2003)]{djg03}		
Djorgovski, S.G., et al., in preparation

\bibitem[Ellison et al.\ (2001)]{ellison01}	
Ellison, S.L., Pettini, M., Steidel, C.C., \& Shapely, A.E. \apj, 549, 770

\bibitem[Ellison et al.\ (2001b)]{ellison01b}    
Ellison, S.L., Yan, L., Hook, I.M., Pettini, M., Wall, J.V.,
\& Shaver, P. 2001, \aap, 379, 393

\bibitem[Junkkarinen et al.\ (2003)]{jnk03}     
Junkkarinen, V.T., et al.\ 2003, in preparation

\bibitem[Lanzetta et al.\ (1995)]{lzwt95}   	
Lanzetta, K. M., Wolfe, A. M.,\&  Turnshek 1995, \apj, 440, 435

\bibitem[Ledoux, Bergeron, \& Petitjean (2002)]{ledoux02a}  
Ledoux, C., Bergeron, J., \& Petitjean, P. 2002, \aap, 385 802

\bibitem[Ledoux, Srianand, \& Petitjean (2002)]{ledoux02b}  
Ledoux, C., Srianand, R. \& Petitjean, P. 2002, \aap, 392, 781

\bibitem[Ledoux, Petitjean, \& Srianand (2003)]{ledoux03}  
Ledoux, C., Petitjean, P., \& Srianand, R. 2003, \aap, submitted (astro-ph/0302582)

\bibitem[Lopez et al.\ (1999)]{lopez99} 	
Lopez, S., Reimers, D., Rauch, M., Sargent, W.L.W., \& Smette, A.
1999, \apj, 513, 598

\bibitem[Lopez et al.\ (2002)]{lopez02}  	
Lopez, S., Reimers, D., D'Odorico, S., \& Prochaska, J.X. 2002, 
\aap, 385, 778

\bibitem[Lu et al.\ (1996)]{lu96}		
Lu, L., Sargent, W.L.W., Barlow, T.A.,
Churchill, C.W., \& Vogt, S. 1996, \apjsupp, 107, 475

\bibitem[Lu et al.\ (1999)]{lu99}
Lu, L., Sargent, W.L.W., \& Barlow, T.A. 1999, in 
{\em Highly Redshifted Radio Lines} ed.\ C.L. Carilli, S.J.E. Radford,
K.M. Menten, and G.I. Langston (San Fransisco: BookCrafters Inc.), p.132
(astro-ph/9711298)  

\bibitem[Mathlin et al.\ (2001)]{mathlin01}		
Mathlin, G.P., Baker, A.C., Churches, D.K., \& Edmunds, M.G. 2001, \mnras,
321, 743

\bibitem[Meyer et al.\ (1995)]{meyer95}
Meyer, D.M., Lanzetta, K.M., and Wolfe, A.M. 1995, \apj, 451, L13

\bibitem[Mirabal et al.\ (2002)]{mirabal02}
Mirabal, N., et al. 2002, ApJ, 578, 818

\bibitem[Molaro et al.\ (2000)]{molaro00}	
Molaro, P., Bonifacio, P., Centuri${\rm \acute o}$n, M.,
D'Odorico, S., Vladilo, G., Santin, P., \& Di Marcantonio, P. 2000,
\apj, 541, 54

\bibitem[Outram et al.\ (2001)]{outram01}    
Outram, P.J., Smith, R.J., Shanks, T., Boyle, B.J., 
Croom, S.M., Loaring, N.S., \& Miller, L. 2001, \mnras, 328, 805

\bibitem[Pei, Fall, \& Bechtold (1991)]{pei91}    
Pei, Y.C., Fall, S.M., \& Bechtold, J. 1991, \apj 378, 6

\bibitem[Pei \& Fall (1995)]{pei95}   
Pei, Y.C. \& Fall, S.M. 1995, \apj, 454, 69

\bibitem[Pei, Fall, \& Hauser (1999)]{pei99}   
Pei, Y.C., Fall, S.M., \& Hauser, M.G. 1999, \apj, 522, 604

\bibitem[P\'eroux et al.\ (2001)]{peroux01}	
P\'eroux, C., Storrie-Lombardi, L.J., McMahon, R.G., Irwin, M., \&
Hook, I.M.  2001, \aj, 121, 1799

\bibitem[P\'eroux et al.\ (2002)]{peroux02}	
P\'eroux, C., Petitjean, P., Aracil, B., \& 
Srianand, R. 2002, New Astr., 7, 577

\bibitem[P\'eroux et al.\ (2003)]{peroux03}	
P\'eroux, C., et al.\ 2003, \aap, submitted

\bibitem[Petitjean et al.\ (2000)]{ptj00}     
Petitjean, P., Srianand, R., \& Ledoux, C. 2000, \aap, 364, 26L

\bibitem[Petitjean et al.\ (2002)]{petit02}     
Petitjean, P., Srianand, R., \& Ledoux, C. 2002, \aap, 332, 383

\bibitem[Pettini et al.\ (1994)]{pettini94}         
Pettini, M., Smith, L. J., Hunstead, R. W., and King,
D. L. 1994, \apj, 426, 79

\bibitem[Pettini et al.\ (1995)]{pettini95}		
Pettini, M., Lipman, K., \& Hunstead, R.W. 1995, \apj, 451, 100

\bibitem[Pettini et al.\ (1999)]{pettini99}		
Pettini, M., Ellison, S.L., Steidel, C.C., \& Bowen, D.V. 1999,
\apj, 510, 576

\bibitem[Pettini et al.\ (2000)]{ptt00}		
Pettini, M., Ellison, S.L., Steidel, C.C., Shapely, A.L., \& Bowden, D.V.
2000, \apj, 532, 65

\bibitem[Pettini et al.\ (2002)]{pettini02}	
Pettini, M., Ellison, S.L., Bergeron, J., \& Petitjean, P. 2002,
\aap, 391, 21

\bibitem[Prochaska (2003a)]{pro03a}  	
Prochaska, J.X. 2003a, \apj, 582, 49

\bibitem[Prochaska (2003b)]{pro03}	
Prochaska, J.X. 2003b, ``Carnegie Symposium: Abundances'', in press

\bibitem[Prochaska, Gawiser, \& Wolfe (2001)]{pgw01}  
Prochaska, J.X., Gawiser, E., \& Wolfe, A.M. 2001, \apj, 552, 99 (PGW01)

\bibitem[Prochaska, Howk, \& Wolfe (2003)]{phw03}  
Prochaska, J.X., Howk, J.C., \& Wolfe, A.M. 2003, \nat, in press

\bibitem[Prochaska \& Wolfe (1996)]{pw96}       
Prochaska, J. X. \& Wolfe, A. M. 1996, \apj, 470, 403

\bibitem[Prochaska \& Wolfe (1997a)]{pw97}		
Prochaska, J. X. \& Wolfe, A. M. 1997, \apj, 474, 140

\bibitem[Prochaska \& Wolfe (1999)]{pw99}
Prochaska, J. X. \& Wolfe, A. M. 1999, \apjs, 121, 369 
 
\bibitem[Prochaska \& Wolfe (2000)]{pw00}	
Prochaska, J.X. \& Wolfe, A.M., 2000, \apj, 533, L5

\bibitem[Prochaska \& Wolfe (2002)]{pw02}	
Prochaska, J.X. \& Wolfe, A.M. 2002, \apj, 566, 68

\bibitem[Prochaska et al.\ (2001)]{pro01}  
Prochaska, J.X., Wolfe, A.M., Tytler, D., Burles, S.M., Cooke, J.,
Gawiser, E., Kirkman, D., O'Meara, J.M., \& Storrie-Lombardi, L.
2001, \apjs, 137, 21

\bibitem[Prochaska et al.\ (2002)]{pro02}  	
Prochaska, J.X., Howk, J.C., O'Meara, J.M., Tytler, D., 
Wolfe, A.M., Kirkman, D., Lubin, D., \& Suzuki, N. 2002, \apj, 571, 693

\bibitem[Prochaska et al.\ (2003a)]{p03a}  
Prochaska, J.X., Gawiser, E., Wolfe, A.M., Cooke, J., \& Gelino, D.
2003a, \apjs, in press

\bibitem[Prochaska et al.\ (2003b)]{pro03b}  
Prochaska, J.X., Castro, S., Djorgovski, S.G. 2003b, \apjs, in press

\bibitem[Prochaska et al.\ (2003c)]{pro03c}  
Prochaska, J.X., et al.\ 2003c, in preparation

\bibitem[Rao \& Turnshek (2000)]{rao00}		
Rao, S.M. \& Turnshek, D.A. 2000, \apjs, 130, 1

\bibitem[Rosenberg \& Schneider (2003)]{rosen03}   
Rosenberg, J.L. \& Schneider, S.E. 2003, \apj, 585, 256

\bibitem[Savaglio, Fall, \& Fiore (2003)]{savaglio03}
Savaglio, S., Fall, S.M., \& Fiore, F. 2003, ApJ, 585, 638

\bibitem[Sheinis et al.\ (2002)]{sheinis02}		
Sheinis, A.I., Miller, J., Bigelow, B., Bolte, M., Epps, H.,
Kibrick, R., Radovan, M., \& Sutin, B. 2002, \pasp, 114, 851

\bibitem[Somerville, Primack, \& Faber (2001)]{somerville01}  
Somerville, R. S., Primack, J. R., \& Faber, S.M. 2001, \mnras, 320, 504

\bibitem[Songaila (2001)]{songaila01}		
Songaila, A. 2001, \apj, 561, 153L

\bibitem[Songaila \& Cowie (2002)]{songaila02}		
Songaila, A. \& Cowie, L.L. 2002, \aj, 123, 2183

\bibitem[Srianand, Petitjean, \& Ledoux (2000)]{srianand00}  
Srianand, R., Petitjean, P., \& Ledoux C. 2000, Nature, 408, 931

\bibitem[Storrie-Lombardi and Wolfe (2000)]{storrie00}  
Storrie-Lombardi, L.J. \& Wolfe, A.M. 2000, \apj, 543, 552

\bibitem[Tissera et al.\ (2001)]{tissera01}     
Tissera, P.B., Lambas, D.G., Mosconi, M.B., \& Cora, S. 2001, \apj, 557, 527

\bibitem[Vladilo et al.\ (2000)]{vladilo00}
Vladilo, G., Bonifacio, P., Centruri\'on, M., \& Molaro, P. 2000,
\apj, 543, 24

\bibitem[Vogt et al.\ (1994)]{vogt94}		
Vogt, S.S., Allen, S.L., Bigelow, B.C., Bresee, L., Brown, B., et al.\ 1994,
SPIE, 2198, 362

\bibitem[Wolfe, Prochaska, \& Gawiser (2003)]{wpg03}	
Wolfe, A. M., Prochaska, J.X., \& Gawiser, E. 2003, \apj, in press

\bibitem[Wolfe, Gawiser, \& Prochaska (2003)]{wgp03}	
Wolfe, A. M., Gawiser, E., \& Prochaska, J.X. 2003, \apj, in press

\bibitem[Wolfe et al.\ (1986)]{wolfe86}
Wolfe, A.M., Turnshek, D.A., Smith, H.E., \& Cohen, R.D.
1986, \apjs, 61, 249

\bibitem[Wolfe et al.\ (1994)]{wolfe94}       	
Wolfe, A. M., Fan, X-M., Tytler, D., Vogt, S. S., Keane, M. J.,
\&  and Lanzetta, K. M. 1994, \apj, 435, L101

\bibitem[Wolfe et al.\ (1995)]{wolfe95}		
Wolfe, A. M., Lanzetta, K. M., Foltz, C. B., and
Chaffee, F. H. 1995, \apj, 454, 698

\end{thebibliography}
\end{document}